\documentclass[12pt,a4paper]{article}
\pdfoutput=1

\usepackage{amsmath,amssymb}
\usepackage[bookmarks=true]{hyperref}
\usepackage[nosort]{cite}
\usepackage[bulletsep]{collect}
\def\gfxon{\usepackage[final]{graphicx}}

\gfxon

\makeatletter
\let\old@startsection=\@startsection
\renewcommand{\@startsection}[6]{\old@startsection{#1}{#2}{#3}{#4}{#5}{#6\mathversion{bold}}}
\makeatother

\makeatletter \@addtoreset{equation}{section} \makeatother

\makeatletter
\let\old@makecaption=\@makecaption
\def\@makecaption{\small\old@makecaption}
\makeatother

\makeatletter
\newcommand{\ellSN}{\mathop{\operator@font sn}\nolimits}
\newcommand{\ellCN}{\mathop{\operator@font cn}\nolimits}
\newcommand{\ellDN}{\mathop{\operator@font dn}\nolimits}
\newcommand{\ellAM}{\mathop{\operator@font am}\nolimits}
\newcommand{\ellK}{\mathop{\smash{\operator@font K}\vphantom{a}}\nolimits}
\newcommand{\ellE}{\mathop{\smash{\operator@font E}\vphantom{a}}\nolimits}
\makeatother

\ifx\genfrac\sdflkaj

\else

\fi
\newcommand{\sfrac}[2]{{\textstyle\frac{#1}{#2}}}
\newcommand{\half}{\sfrac{1}{2}}

\newcommand{\beq}{\begin{equation}}
\newcommand{\eeq}{\end{equation}}

\def\[{\begin{equation}}
\def\]{\end{equation}}
\def\<{\begin{eqnarray}}
\def\>{\end{eqnarray}}

\makeatletter
\def\mr@ignsp#1 {\ifx\:#1\@empty\else #1\expandafter\mr@ignsp\fi}%
\newcommand{\multiref}[1]{\begingroup
\xdef\mr@no@sparg{\expandafter\mr@ignsp#1 \: }%
\def\mr@comma{}%
\@for\mr@refs:=\mr@no@sparg\do{\mr@comma\def\mr@comma{,}\ref{\mr@refs}}%
\endgroup}
\makeatother

\newcommand{\hypref}[2]{\ifx\href\asklfhas #2\else\href{#1}{#2}\fi}

\newcommand{\secref}[1]{Sec.~\multiref{#1}}

\newcommand{\figref}[1]{Fig.~\multiref{#1}}
\renewcommand{\eqref}[1]{(\multiref{#1})}

\ifx\href\asklfhas\newcommand{\href}[2]{#2}\fi
\newcommand{\arxivno}[1]{\href{http://arxiv.org/abs/#1}{#1}}

\begin{document}

\begin{flushright}\footnotesize
\texttt{ArXiv:\arxivno{0909.2551}}\\
\texttt{ITEP-TH 40/09}\\
\texttt{FTPI-MINN-09/33}\\
\texttt{UMN-TH-2813/09}\\
\vspace{0.5cm}
\end{flushright}
\vspace{0.3cm}

\renewcommand{\thefootnote}{\arabic{footnote}}
\setcounter{footnote}{0}

\begin{center}%
{\Large\textbf{\mathversion{bold}
Wilson Loops in Gravity Duals of Lifshitz-like Theories}
\par}

\vspace{1cm}%

\textsc{Peter Koroteev$^{\#}$ and  A.~V.~Zayakin$^{\& \dagger}$}

\vspace{10mm}

\textit{$^{\#}$University of Minnesota, School of Physics and Astronomy\\%
116 Church Street S.E. Minneapolis, MN 55455, USA}

\vspace{2mm}

\textit{$^{\&}$Fakult\"at f\"{u}r Physik der Ludwig-Maximillians-Universit\"{a}t M\"{u}nchen \\
und Maier-Leibniz-Laboratory, Am Coulombwall 1, 85748 Garching, Germany}

\vspace{2mm}

\textit{$^\dagger$ Institute of Theoretical and Experimental Physics,\\
 Moscow 117218, Russia}

\vspace{7mm}

\thispagestyle{empty}

\texttt{koroteev@physics.umn.edu,\,Andrey.Zayakin@physik.lmu.de}

\par\vspace{1cm}

\vfill

\textbf{Abstract}\vspace{5mm}

\begin{minipage}{12.7cm}
We calculate Wilson loops on boundaries of Lifshitz-like dual backgrounds with different scaling parameters, assuming existence of a field theory dual to string theory in the bulk. We consider scaling parameters to be variable quantities which are subject to cosmological evolution. It is observed that there are discontinuities in the classical string action at some values of the scaling parameters. 
\end{minipage}

\vspace*{\fill}

\end{center}

\newpage

\section*{Introduction}\label{sec:Intro}

Interest in theories with dynamical scaling and broken Lorentz invariance\footnote{In this paper we shall use ``Lorentz invariance'' and ``conformal invariance'' equivalently.} has recently been warmed up, driven by a number of observations. First, cosmological models with broken Lorentz invariance \cite{Libanov:2005nv, Libanov:2005vu, Libanov:2005yf, Libanov:2007mq} were shown to be useful tools for description of the early Universe; those models including extra dimensions \cite{Dubovsky:2001fj, Koroteev:2007yp, Koroteev:2009xd} appear to possess a very unusual spectrum of field perturbations. Second, a holography dual programme for Lifshitz-like theories has been outlined in \cite{Kachru:2008yh} which uncovered many questions in gauge/gravity duality and possibilities to generalize it to a nonconformal case. Third, Horava's proposal on generalizations of Einstein gravity to spacetimes with nontrivial dynamical scaling \cite{Horava:2009if, Horava:2009uw} revealed a new prospective upon connection of gravity to string theory. There are, however, some obstacles in this direction (see \cite{Blas:2009yd} and others).

The AdS/CFT correspondence \cite{Maldacena:1997re, Gubser:1998bc} has been studied to a very high extent, and one of the reasons for this success is a large number of symmetries which are manifest on both sides of the duality. However, duality in the nonconformal case still requires further clarification. So far, only a few non-local objects in non-conformal gauge theories have been considered, e.g. Wilson loops in~\cite{PandoZayas:2008fw}; it is a natural willingness to extend these considerations to a wider class of geometries. This is the goal of the current paper -- we calculate Wilson loops on the holographic boundary of spaces with broken conformal invariance with a Lifshitz-like metric introduced in \cite{Koroteev:2007yp, Koroteev:2009xd}.

The information one can get from a Wilson loop is quite rich. By analyzing Wilson (or Polyakov) loops, one can justify about renormalization group flows, color potential, confinement property and other important features\cite{Wilson:1974sk,West:1982bt}. We know a number of fascinating properties of some BPS and non-BPS loops in the Lorentz-invariant theory, namely, the possibility to obtain a Wilson loop at strong coupling from both parts of duality --- by ladder resummation from field theory \cite{Erickson:1999qv,
Erickson:2000af}, and by string area calculation from AdS/CFT correspondence \cite{Berenstein:1998ij}. According to Maldacena conjecture, to calculate a Wilson loop on the boundary, one should consider a classic worldsheet of a string which has the loop as its boundary \cite{Maldacena:1998im,Rey:1998bq}.

Let us keep in mind that the correspondence between calculations of Wilson loops on the boundary and the area of the minimal surface in the bulk spanned over the loop is supposed to be true for the conventional AdS/CFT duality \cite{Maldacena:1998im} where conformal invariance is unbroken. Our calculation is performed in the absence of the conformal symmetry in the bulk, therefore there are no guarantees that the above matching will take place. The legitimacy of our calculation can be questioned together with other holographic calculations which were performed in theories with a nontrivial dynamical scaling. We are not going to address this question in the current work; rather, we want to show that once Maldacena conjecture can be generalized to backgrounds with broken Lorentz invariance, the calculation we do may uncover some interesting physical phenomena.

In this work we calculate Wilson loops which consist of two parallel long lines located apart from each other at some fixed distance. In the first case these lines are parallel to the time axis and in the second case they are parallel to the coordinate axis on the boundary. We then focus on the dependence of expectation values of the two Wilson loops we calculate on the scaling parameters (or anisotropy/Lorentz violating parameters). The difference between the two calculations shows how nontrivial dynamical scaling
\[
t\to \lambda^\xi t\,,\quad x\to \lambda^\zeta x\,,
\]
brings anisotropy in the bulk theory.  We observe that dependence of the above Wilson loops on $\xi$ and $\zeta$ possess discontinuities.

The paper is organized as follows. In \secref{sec:LIVbackgrounds} we summarize some facts about backgrounds with broken Lorentz invariance we need for calculation of Wilson loops. Then in \secref{sec:Wilson} we present calculations of space-like and time-like Wilson loops located on the boundary of backgrounds discussed in previous section. In the end we give conclusions and pose some questions.

\section{Backgrounds with Broken Lorentz Invariance}\label{sec:LIVbackgrounds}

In \cite{Koroteev:2009xd} a classification of static backgrounds with broken Lorentz invariance was given. Metric in question has the following form
\[\label{eq:MetricKL}
ds^2 = L^2\left(-r^{2\xi}dt^2+r^{2\zeta}d\textbf{x}^2+\frac{dr^2}{r^2}\right)
\]
where $(t,\textbf{x})$ are branelike coordinates and $r$ is the extra dimensional (holographic) coordinate. It can be easily shown that there is no coordinate transformation which brings this metric to the conformal Minkowski metric unless $\xi=\zeta$. This implies that Lorentz invariance is broken. In what follows we shall consider only $1+1+1$  dimensional spaces of type \eqref{eq:MetricKL} (i.e. $\textbf{x}$ is one-dimensional). However, our considerations can be straightforwardly applied for $1+d+1$ dimensional spacetimes as well.

The solution \eqref{eq:MetricKL} was first obtained in \cite{Koroteev:2007yp} and the parameters $\xi$ and $\zeta$ are triggered by anisotropy of the bulk matter. Slightly later in \cite{Kachru:2008yh} a microscopic description of the model has been discovered (see \cite{Gordeli:2009vh} where an exhaustive comparison of the two above solutions is made).

Let us mention that the whole family of spaces \eqref{eq:MetricKL} have constant negative curvature which reduces many computational difficulties in gravity and field theory calculations. In principle one may think of more sophisticated metric, but, as is argued in \cite{Koroteev:2009xd}, it will probably not bring too much new physics.

\begin{figure}[!ht]
\unitlength=1mm
\begin{picture}(110,110)(-30,-5)
\includegraphics[height = 10cm, width=11.5cm]{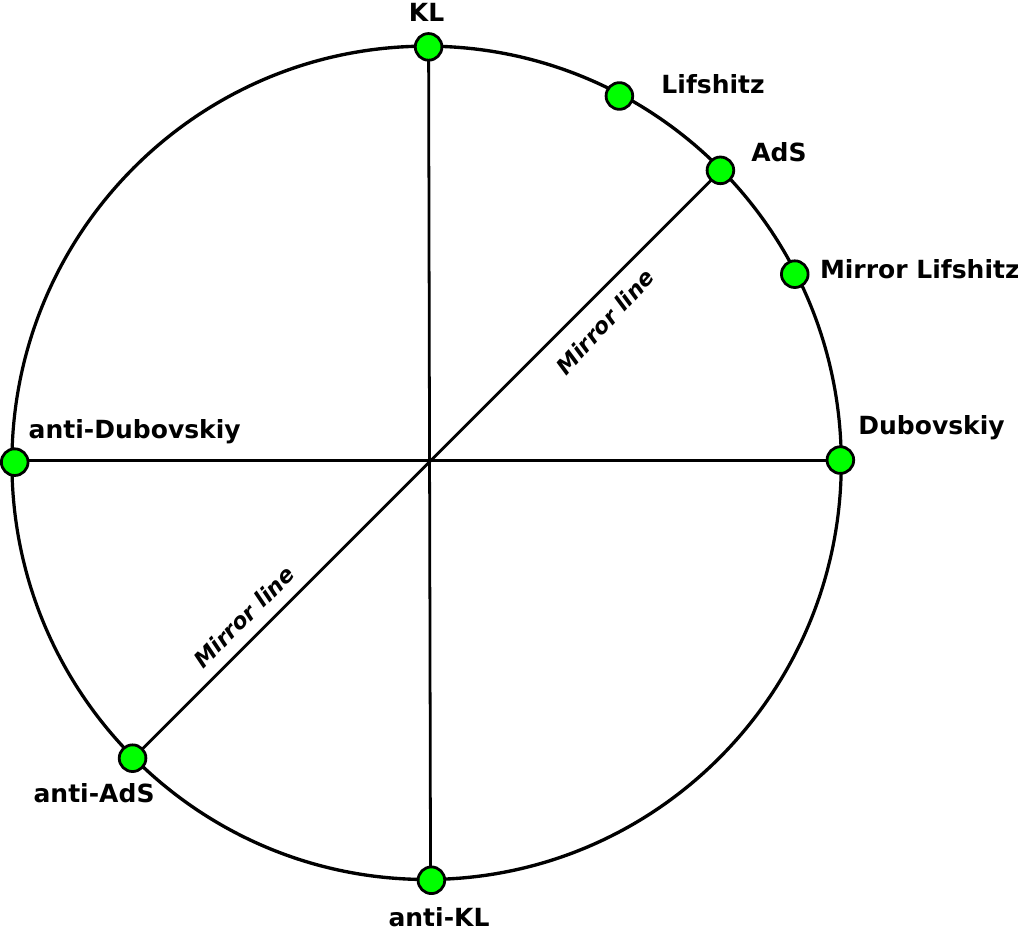}
\put(-26,53){$\xi$}
\put(-73,89){$\zeta$}
\end{picture}
\caption{The parameter space. All known exactly solvable models are schematically shown on a circle implicitly assuming $\xi^2+\zeta^2=1$ which is however not necessary. Mirror line $\xi=\zeta$ illustrates mirror transformations as reflections with respect to it. Thus Lifshitz model has its mirror dual and the KL model is mirror dual to the Dubovsky model. Models with signs of $\xi$ and $\zeta$ flipped are called anti-models.}
\label{fig:xizeta}
\end{figure}

The space of metrics of the form \eqref{eq:MetricKL} parameterized by $\xi,\zeta$ is shown in \figref{fig:xizeta}. Evidently spaces parameterized by \eqref{eq:MetricKL} comprise a one-dimensional family and keeping two parameters $\xi$ and $\zeta$ is a redundancy\footnote{One can impose some constraint on them e.g. $\xi^2+\zeta^2=1$}, nevertheless it enables us to monitor scaling properties of $t$ and $x$ independently.
The model at the point $A$ was elaborated in \cite{Koroteev:2009xd}, the model corresponding to the point $B$ was discussed in \cite{Dubovsky:2001fj}; the other known exact solutions include Lifshitz model $(\xi=2,\zeta=1)$ and its mirror dual $(\xi=1,\zeta=2)$. Surely, for $\xi=\zeta=1$ there exists the AdS solution.

An interesting behavior of the metric \eqref{eq:MetricKL} is observed when one of the parameters $\xi$ or $\zeta$ or both become negative. Indeed, for positive $\xi$ and $\zeta$ the UV boundary is located at $r\to \infty$ and is two dimensional. However, if, say, $\zeta<0$ then the $x$ direction shrinks as we approach the UV boundary and the boundary becomes effectively one-dimensional. In the opposite IR limit $r\to 0$ due to $r^{2\zeta}\to\infty$, $x$-direction expands. In the next section we shall calculate two Wilson loops located at the UV boundary and investigate how these loops depend on scaling parameters $\xi$ and $\zeta$. It is natural to expect that something interesting should happen when these parameters passes through the origin. Further calculations confirm this conjecture.

\section{Wilson Loops}\label{sec:Wilson}

We start with the metric \eqref{eq:MetricKL} and work with the Nambu--Goto action
\[\label{eq:NGAction}
S=\int \sqrt{\det_{\alpha,\beta} \partial_\alpha X^A \partial_\beta X^B g_{AB}} d\sigma d\tau,
\]
where $X^A$ are embedding coordinates, $\tau,\sigma$ are worldsheet coordinates, and $g_{AB}$ is target space metric \eqref{eq:MetricKL}.

\begin{figure}[!h]
\begin{center}
\includegraphics[height = 5cm, width=9cm]{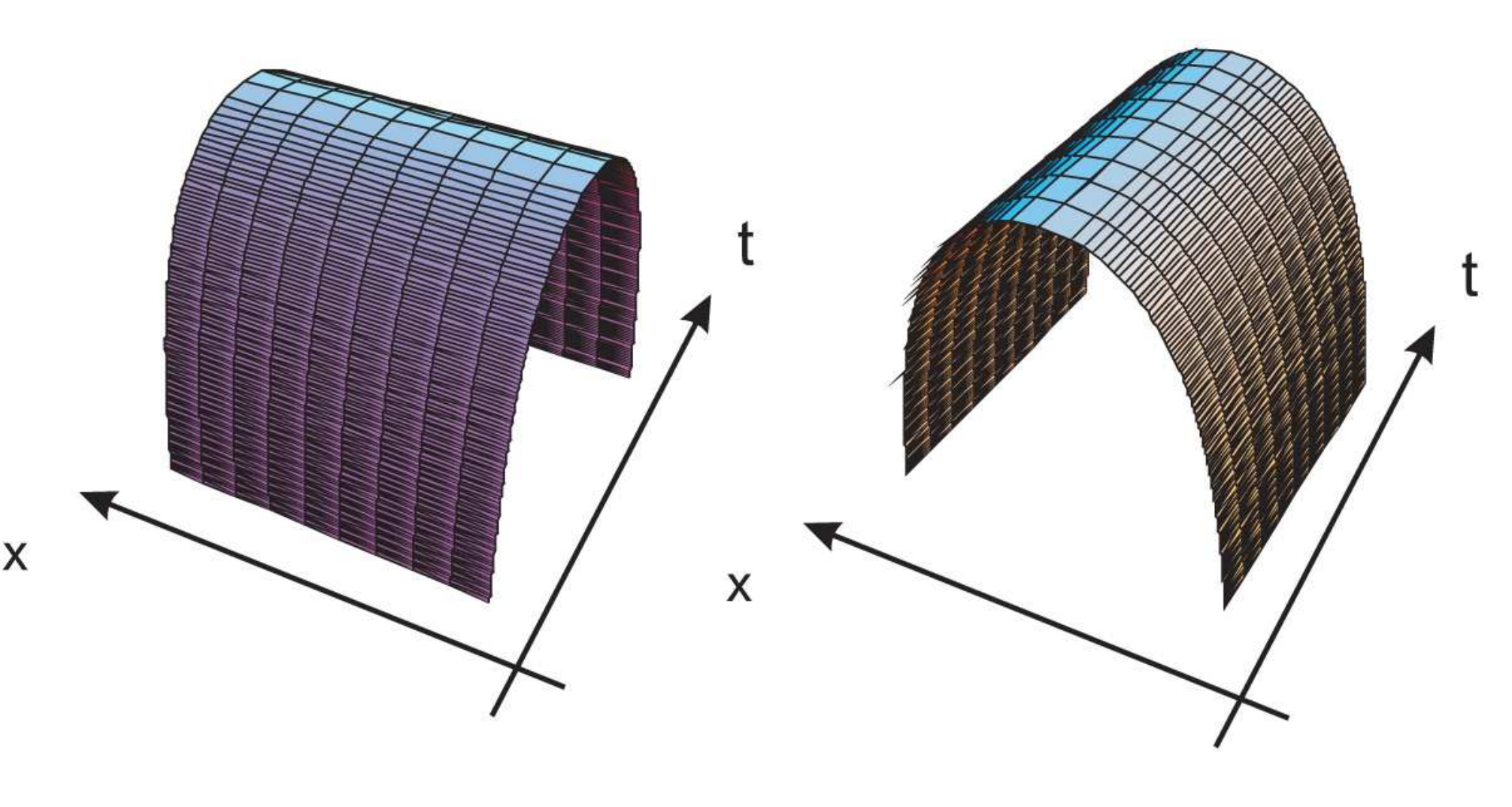}
\caption{Geometry of space-like (left plot) and time-like (right plot) Wilson loops.}
\label{fig:RectWilsonLoopsGeom}
\end{center}
\end{figure}

We calculate each of the two loops depicted in \figref{fig:RectWilsonLoopsGeom} --- a time-like and a space-like loop.
Calculations in these cases can be done analytically for any value of the scaling parameters $\zeta,\xi$. The goal is to study the dependence of
classical action \eqref{eq:NGAction} on $\zeta,\xi$.

The meaning of the time-like loop calculation is clear -- it represents interaction of two static quarks located at some distance apart from each other. However, the space-like loop corresponding to instantaneous propagation does not have any immediate interpretation. A space like loop in a theory with metric \eqref{eq:MetricKL}
is equal to the time-like loop in the mirror theory with metric\footnote{From now on we shall use Euclidian signature}
\[\label{eq:MetricKLMirror}
ds^2 = L^2\left(r^{2\zeta}dt^2+r^{2\xi}dx^2+\frac{dr^2}{r^2}\right)\,.
\]
Therefore, we may calculate the both Wilson loops for a particular metric, e.g. \eqref{eq:MetricKL} and interpret the result for the space-like loop as the answer for the time-like loop in the mirror theory \eqref{eq:MetricKLMirror}.

For the surfaces we consider here the embedding function has the following form
\[
X^A=\begin{pmatrix}\tau & \sigma & r(\tau,\sigma)\end{pmatrix}\,,
\]
where worldsheet coordinates $\tau,\sigma$ are trivially mapped onto $(t,x)$ plane at some constant value of $r$, e.g. onto the UV boundary. The action \eqref{eq:NGAction} for the above embedding reads
\[\label{eq:NGActionSpace}
S=\int d\sigma d\tau \sqrt{r^{2\xi+2\zeta}+r^{2(\xi-1)}r'^2+r^{2(\zeta-1)}\dot{r}^2}\,.
\]
This action remains the same if we interchange
\[\label{eq:MirrorTransformations}
\xi\longleftrightarrow \zeta, \quad t \longleftrightarrow x
\]
simultaneously. However, a different functional can be obtained if only one of the above transformations is performed. We shall refer to either of those transformations as mirror transformations and shall use them to obtain mirror solutions.

In the current paper we shall calculate only rectangular loops, so dependence on one of the parameters $\tau$ or $\sigma$ is trivial for these loops. Let us first consider the time-like Wilson loop, then the Nambu-Goto action has the following form
\[\label{eq:NGTime}
S=\int d\sigma d\tau \sqrt{r^{2\xi+2\zeta}+r^{2(\xi-1)}r'^2}\,,
\]
and time derivative $\dot{r}$ vanishes. Translational invariance along $t$ direction enables us to find the following conserved charge
\[\label{eq:IntofMotTime}
\mathcal{H}=-\frac{r^{2\xi+2\zeta}}{\sqrt{r^{2\xi+2\zeta}+r^{2(\xi-1)}r'^2}}.
\]
It allows us to obtain the solution in quadratures
\[\label{eq:SolutionTime}
\int\limits_{0}^R d\sigma=r_0^{-\zeta}\int\limits_{u_0}^{u_1} \frac{du}{u^{1+\zeta}\sqrt{u^{2\xi+2\zeta}-1}},
\]
where $u=r/r_0$ and $r_0^{\xi+\zeta}=\mathcal{H}$. The integration limits in the r.h.s. depend on the scaling parameters. Evidently in order to have a finite real-valued solution there will be different regimes for $\xi+2\zeta>0$ and $\xi+2\zeta<0$. Let us consider these cases separately.

\paragraph{UV configuration: $\xi+2\zeta>0$.}

In this case integration limits in \eqref{eq:SolutionTime} are $u_0=1,u_1=+\infty$ and the loop is located at the UV boundary $u=u_1$. Performing the integration we obtain the boundary conditions
\[\label{eq:r0spacelike}
R=r_0^{-\zeta}\int\limits_1^{+\infty}\frac{du}{u^{1+\zeta}\sqrt{u^{2\xi+2\zeta}-1}}
=r_0^{-\zeta}\frac{\sqrt{\pi}\Gamma
\left(\frac{\xi+2\zeta}{2\xi+2\zeta}\right)}{\zeta\Gamma\left(\frac{\zeta}{2\xi+2\zeta}\right)}\,.
\]
Making use of \eqref{eq:r0spacelike}, on-shell action \eqref{eq:NGActionSpace} becomes
\[
S=2T \cdot2r_0^\xi\int\limits_1^\infty du \frac{u^{2\xi+\zeta-1}}{\sqrt{u^{2\xi+2\zeta}-1}}\,,
\]
where $T\gg R$ is the length of Wilson lines along $t$ direction. Here reguralization of the above integral is needed. Recall that in order to obtain physical result for the calculation of the area we need to subtract the contribution of free
quarks i.e. twice the area of planes $x=\pm R$ which span $t$ and $r$ directions. Namely,
\[\label{eq:RegAction}
S_{\text{reg}}=S-\frac{4T}{R^{\xi/\zeta}}r_0^\xi\int\limits_0^{+\infty}\frac{du}{u^{1-\xi}}
=\frac{4T}{R^{\xi/\zeta}}r_0^\xi\left(\int\limits_1^{+\infty}\,du\left(\frac{u^{2\xi+\zeta-1}}{\sqrt{u^{2\xi+2\zeta}-1}}-\frac{1}{u^{1-\xi}}\right)-\int\limits_0^1\frac{du}{u^{1-\xi}}\right)\,,
\]
which, due to the last term, requires $\xi>0$. Finally we obtain
\[\label{eq:RegActionSol}
S^{\text{UV}}_{\text{reg}}(\xi,\zeta)=\frac{4T}{R^{\xi/\zeta}}
\frac{\pi^{\frac{\xi+\zeta}{2\zeta}}}{2\xi+2\zeta}\frac{\Gamma\left(-\frac{\xi}{2\xi+2\zeta}\right)}{\Gamma\left(\frac{\zeta}{2\xi+2\zeta}\right)}\left[
\frac{\Gamma\left(\frac{\xi+2\zeta}{2\xi+2\zeta}\right)}{\zeta\Gamma\left(\frac{\zeta}{2\xi+2\zeta}\right)}\right]^{\frac{\xi}{\zeta}}\,.
\]
Note that convergence of the normalization integral in the above calculation restricts the domain of scaling parameters allowing to obtain a finite result. It remains unclear how to interpret the calculation in the region where the normalization integral diverges.

\paragraph{IR configuration: $\xi+2\zeta\leq 0$.}

Here the integration limits in \eqref{eq:SolutionTime} are $u_0=1$ and $u_1=0$ and the loop is located at the IR boundary. Carrying out the calculations analogously to the UV case we obtain the same solution as \eqref{eq:RegActionSol} but with signs of $\xi$ and $\zeta$ flipped. This shows us the the IR configuration is isomorphic to the UV configuration under appropriate mapping of the scaling parameters. Performing renormalization along with \eqref{eq:RegActionSol} we can observe that it is possible only for $\xi<0$ which restricts the domain of allowed scaling parameters. Thus we state that
\[\label{eq:RegActionSolIR}
S^{\text{IR}}_{\text{reg}}(\xi,\zeta)=-S^{\text{UV}}_{\text{reg}}(-\xi,-\zeta)
\]
provided that $\xi+2\zeta<0,\,\xi<0$.

\begin{figure}[!h]
\unitlength=1mm
\begin{picture}(110,110)(-30,-5)
\includegraphics[height = 10cm, width=11.5cm]{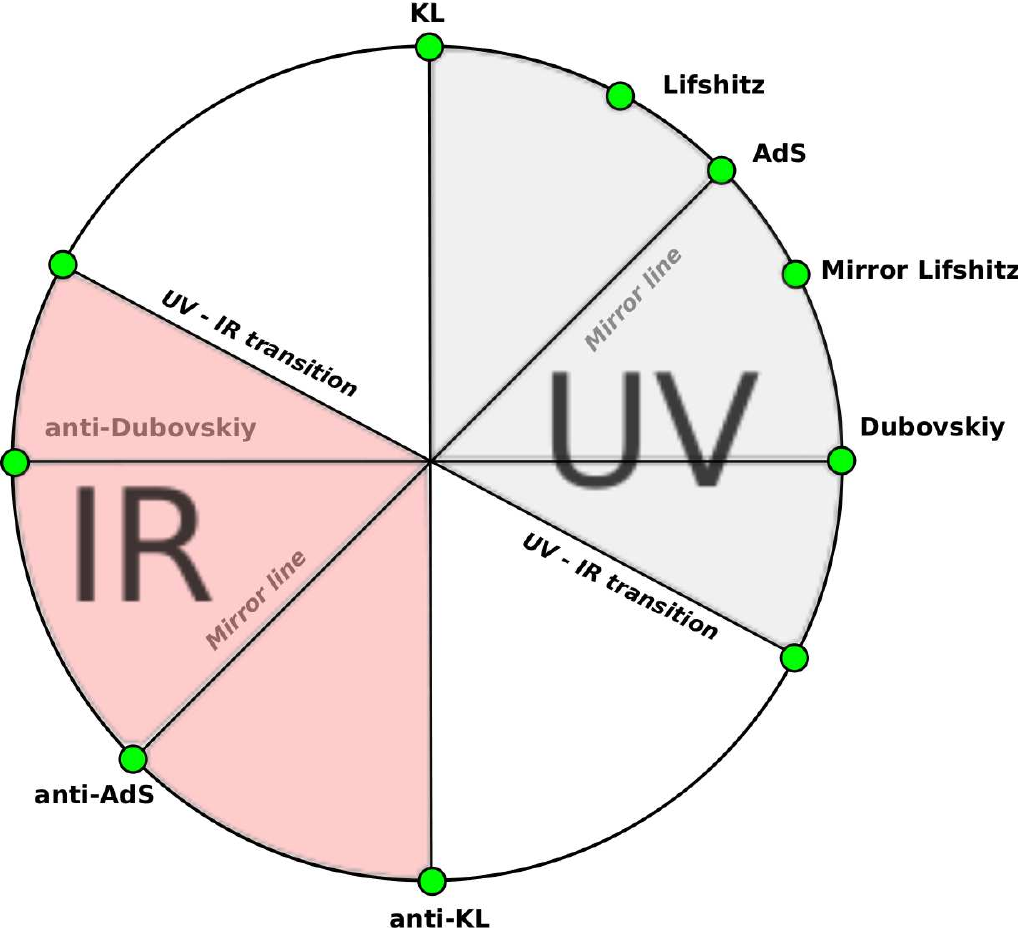}
\put(-26,53){$\xi$}
\put(-73,89){$\zeta$}
\end{picture}
\caption{Regions in the parameter space where renormalized classical action exists. UV and IR configurations are emphasized.}
\label{fig:xizetaUVIR}
\end{figure}

To summarize our understanding of the region where the solution exist we draw a picture similar to \figref{fig:xizeta} but with UV and IR regions marked \figref{fig:xizetaUVIR}
%

\paragraph{Mirror Solutions.}

As was argued above, the transformations \eqref{eq:MirrorTransformations} may lead us to different solutions in mirror metric. Thus if we apply both transformations, the actions \eqref{eq:RegActionSol, eq:RegActionSolIR} are to be interpreted as solutions for a space-like Wilson loop in the mirror metric to \eqref{eq:MetricKLMirror}. In order to find the solutions for the space-like loop in the original metric \eqref{eq:MetricKL} and equivalently for the time-like loop in the mirror metric \eqref{eq:MetricKLMirror} we need to interchange $\xi$ and $\zeta$ in \eqref{eq:RegActionSol, eq:RegActionSolIR}.

\begin{figure}[!h]
\begin{center}
\includegraphics[height = 5cm, width=7cm]{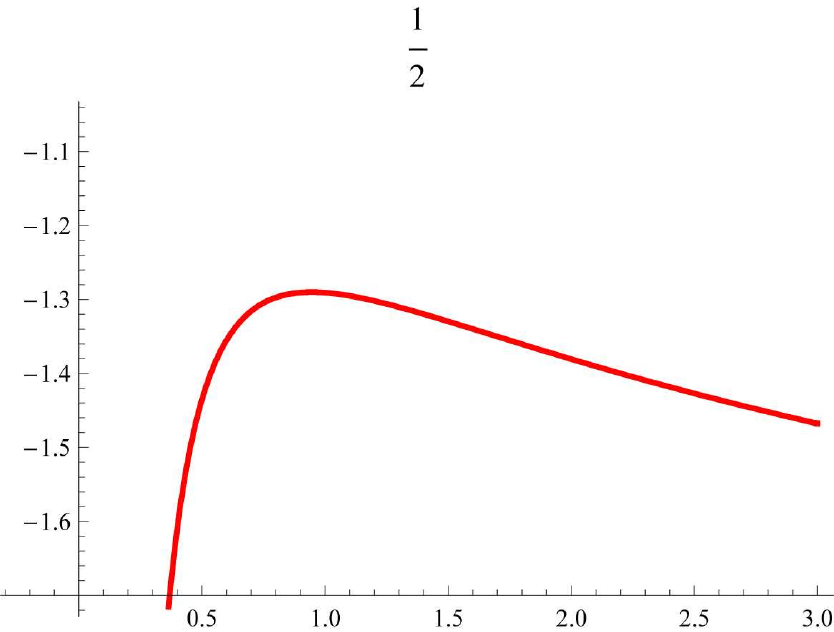} \quad \includegraphics[height = 5cm, width=7cm]{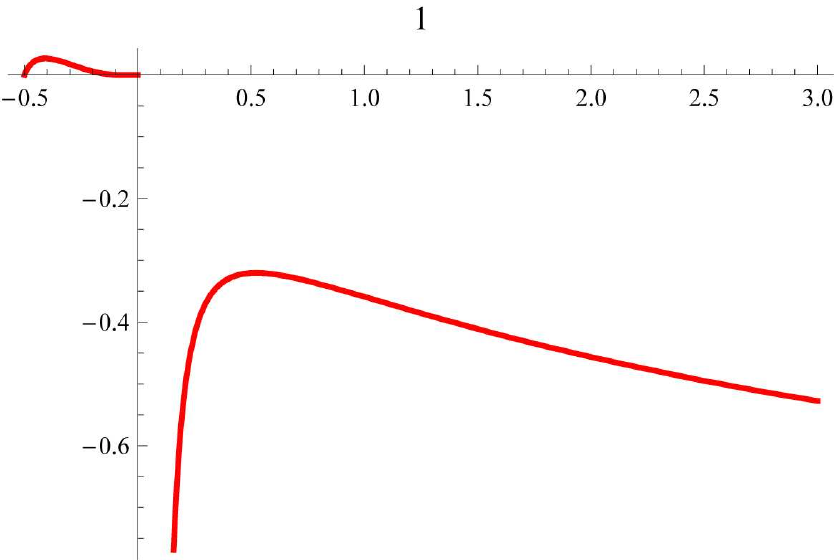} \quad \includegraphics[height = 5cm, width=7cm]{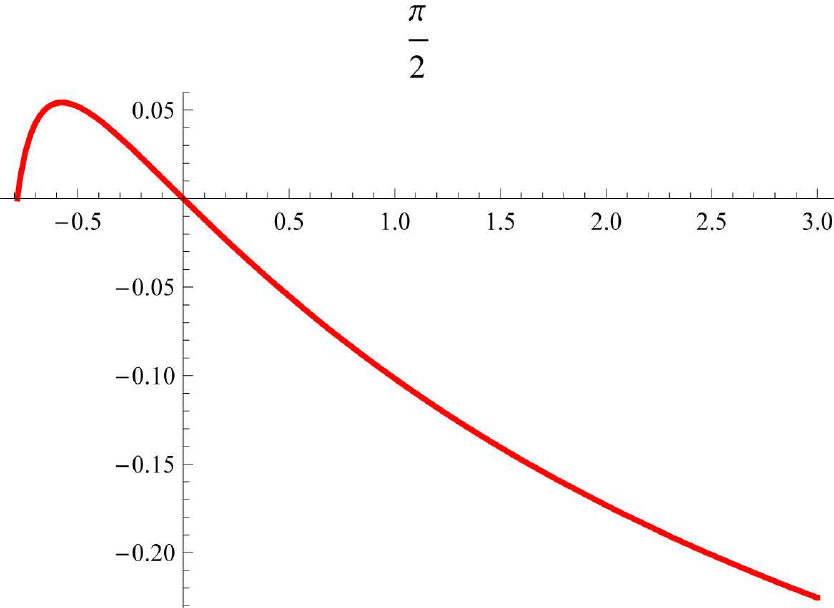} \quad \includegraphics[height = 5cm, width=7cm]{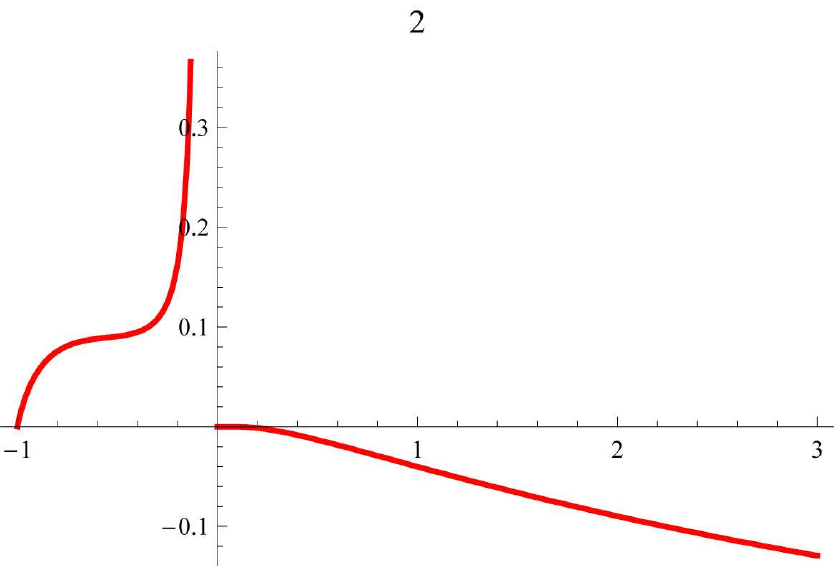}
\caption{Dependence of classical action on $\zeta$ for $\xi=\half,1,\sfrac{\pi}{2},2$ (top left, top right, bottom left, bottom right plots accordingly). All plots correspond to the UV configuration. The solution does not exist where no plots are drawn due to renormalizability problems. An interesting phenomena occurs when $\xi$ passes the point $\sfrac{\pi}{2}$ where two branches merge and then again separate. Only at this point the solution becomes continuous and it remains discontinuous otherwise.}
\label{fig:RectWilsonLoops}
\end{center}
\end{figure}

Various plots showing dependence of $S_{\text{reg}}$ the action on the scaling parameters are presented in \figref{fig:RectWilsonLoops}.
%

\section{Conclusions and Outlook}\label{sec:Conclusions}

In this paper we have calculated rectangular time-like and space-like Wilson loops located on the boundary of Lifshitz spacetime upon assuming existence of its holography dual. We investigated dependence of the classical actions on the scaling parameters and observed discontinuities at some values of these parameters. These discontinuities may be related to some phase transitions, caused by the variation of the space-time geometry.

Another novel observation was made concerning the IR configurations. Namely, we showed that due to renormalization issues, not at all values of scaling parameters a solution for a Wilson loop at the UV boundary exists. A natural proposal here is to move to the IR boundary which, due to a nontrivial dynamical scaling, also exists. Then one can pose the same boundary problem but instead for the IR configuration. We explicitly found a solution for the IR configuration. Nevertheless there still exists a region of parameters (nonshaded regions in \figref{fig:xizetaUVIR}) where renormalization by means of infinitely heavy quarks fails to work. It remains unclear how to interpret a calculation of minimal area of the surface spanned over the Wilson loop on either of the boundaries.

We have observed an interesting dependence of Wilson loops on the scaling parameters $\xi,\zeta$, however, one may ask, why does one need to care about $\xi,\zeta$-dependence if each time we study a theory with fixed value of the scaling parameters? Certainly the observed dependence makes sense if $z=\zeta-\xi$ is a dynamical parameter i.e. we deal with some cosmological scenario, and metric depends on time,
for instance, as follows
\[\label{eq:MetricKLevol}
ds^2 = L^2\left(-f(r,t)^{2\xi(t)} dt^2+f(r,t)^{2\zeta(t)}dx^2+\frac{dr^2}{r^2}\right)\,,
\]
or, perhaps, in a more intricate way. Dramatic changes of a theory on the boundary should happen when either $\xi$ or $\zeta$ passes its junction points which were discussed earlier. In a nonabelian gauge theory these junctions may be related to some phase transitions under which interaction of constituent particles is abruptly changed. Here we do not claim our understanding of the details of these transitions, rather, we want to point out their existence.

Thus, one needs to build up a cosmological setup which employs the above metric and solves the corresponding Einstein equations. Recall that existence of such a solution is crucial for some cosmological scenarios with broken Lorentz invariance \cite{Libanov:2005nv, Libanov:2005yf} which use a metric similar to \eqref{eq:MetricKLevol}. By now no such solution has been found, but, hopefully, due to recently discovered microscopic description \cite{Kachru:2008yh}, searching for this solution is within reach.

The analysis performed in this paper should be complemented by calculation of Wilson loops of other configurations as well as interaction of Wilson loops with each other and local operators. Certainly, a very good argument to confirm (or disprove) our findings would be to elaborate the dual gauge theory explicitly, to study a Wilson loop by means of ladder diagram resummation and to observe its $\xi,\zeta$--dependence. Unfortunately, we lack a field-theoretical description of the corresponding dual gauge model so far.

\paragraph{Note added.}

While working on the current paper we were unaware of the result \cite{Danielsson:2009gi} where the time-like rectangular Wilson loop had been calculated. The authors had obtained the dependence of the classical action for the metric with one single scaling parameter (critical exponent). In our contribution we have generalized this result to the wider class of backgrounds with two scaling parameters and showed that dependence on both parameters is important for the classical action (see \figref{fig:RectWilsonLoops}).

\section*{Acknowledgments}

We are grateful to A. Gorsky, A. Monin, M. Shifman, A. Tseytlin, and A. Vainshtein for fruitful discussions during our work on this paper. We also want to thank R. Radpour for reviewing the final version of the manuscript. This work was supported in part by the DOE grant DE-FG02-94ER40823 (P.K.), by the grant 07-01-00526 (A.Z.), and also supported by the DFG Cluster of Excellence MAP (Munich Centre of Advanced Photonics) (A.Z.).

\bibliography{holliv}
\bibliographystyle{nb}

\end{document}